\documentclass[preprint]{vgtc}               

\ifpdf
  \pdfoutput=1\relax                   
  \usepackage{graphicx}                
  \DeclareGraphicsExtensions{.pdf,.png,.jpg,.jpeg} 
\else
  \usepackage{graphicx}                
  \DeclareGraphicsExtensions{.eps}     
\fi%

\graphicspath{{figures/}{pictures/}{images/}{./}} 

\usepackage{microtype}                 
\PassOptionsToPackage{warn}{textcomp}  
\usepackage{textcomp}                  
\usepackage{mathptmx}                  
\usepackage{times}                     
\usepackage{cite}                      
\usepackage{tabu}                      
\usepackage{booktabs}                  

\usepackage{pbox}
\usepackage{makecell}

\newcommand{\cit}[1]{``#1''}
\newcommand{\ignore}[1]{}

\onlineid{0}

\vgtccategory{Research}

\vgtcinsertpkg

\makeatletter
\preprinttext{\parbox{0.9\textwidth}{$\copyright$ \the\year~IEEE. This is the author's version of the article that has been published in Proceedings of the 2022 IEEE Workshop on Evaluation and Beyond --- Methodological Approaches to Visualization (BELIV~'22). The final version of this record is available at: \href{https://doi.org/\vgtc@DOI}{\color{blue}\vgtc@DOI}}}
\makeatother

\ieeedoi{10.1109/BELIV57783.2022.00008}

\hypersetup{
	pdfinfo={
		Title={An Interdisciplinary Perspective on Evaluation and Experimental Design for Visual Text Analytics: Position Paper},
		Author={Kostiantyn Kucher, Nicole Sultanum, Angel Daza, Vasiliki Simaki, Maria Skeppstedt, Barbara Plank, Jean-Daniel Fekete, Narges Mahyar},
		Keywords={Evaluation; User Study; Visual Text Analytics; Text Visualization; Visual Analytics; Information Visualization; Natural Language Processing; Computational Linguistics; Text Mining},
  }
}

\title{An Interdisciplinary Perspective on Evaluation and Experimental Design for Visual Text Analytics: Position Paper}

\author{Kostiantyn Kucher\thanks{e-mail: kostiantyn.kucher@\{liu.se,lnu.se\}}\\%
     \parbox{3.5cm}{\scriptsize \centering Linköping University\\Linnaeus University\medskip}%
\and Nicole Sultanum\thanks{e-mail: nicolebs@cs.toronto.edu}\\%
     \parbox{3.5cm}{\scriptsize \centering University of Toronto}%
\and Angel Daza\thanks{e-mail: j.a.dazaarevalo@vu.nl}\\%
     \parbox{3.5cm}{\scriptsize \centering Vrije Universiteit Amsterdam}%
\and Vasiliki Simaki\thanks{e-mail: vasiliki.simaki@englund.lu.se}\\%
     \parbox{3.5cm}{\scriptsize \centering Lund University}%
\and Maria Skeppstedt\thanks{e-mail: maria.skeppstedt@abm.uu.se}\\%
     \parbox{3.5cm}{\scriptsize \centering CDHU, Department of ALM,\\ Uppsala University}%
\and Barbara Plank\thanks{e-mail: bplank@\{cis.lmu.de,itu.dk\}}\\%
     \parbox{3.5cm}{\scriptsize \centering LMU Munich\\IT University of Copenhagen}%
\and Jean-Daniel Fekete\thanks{e-mail: jean-daniel.fekete@inria.fr}\\%
     \parbox{3.5cm}{\scriptsize \centering Inria\\Université Paris-Saclay}%
\and Narges Mahyar\thanks{e-mail: nmahyar@cs.umass.edu}\\%
     \parbox{3.5cm}{\scriptsize \centering UMass Amherst}%
}

\abstract{Appropriate evaluation and experimental design are fundamental for empirical sciences, particularly in data-driven fields. 
Due to the successes in computational modeling of languages, for instance, research outcomes are having an increasingly immediate impact on end users. 
As the gap in adoption by end users decreases, the need increases to ensure that tools and models developed by the research communities and practitioners are reliable, trustworthy, and supportive of the users in their goals. 
In this position paper, we focus on the issues of evaluating visual text analytics approaches. 
We take an interdisciplinary perspective from the visualization and natural language processing communities, as we argue that the design and validation of visual text analytics include concerns beyond computational or visual/interactive methods on their own. 
We identify four key groups of challenges for evaluating visual text analytics approaches (data ambiguity, experimental design, user trust, and ``big picture'' concerns) and provide suggestions for research opportunities from an interdisciplinary perspective. %
} 

\CCScatlist{
  \CCScatTwelve{Human-centered computing}{Visu\-al\-iza\-tion}{Visualization design and evaluation methods}{}
}

\begin{document}


\firstsection{Introduction}
\label{sec:introduction}

\maketitle

\renewcommand{\figureautorefname}{Figure}
\renewcommand{\sectionautorefname}{Section}
\renewcommand{\subsectionautorefname}{Section}

The research on computational methods for language and specifically text data, in particular, natural language processing (NLP), has achieved impressive advances in the past several decades. 
The more traditional approaches have been complemented and in some cases replaced by data-driven approaches relying on machine learning (ML) and deep learning (DL), resulting in very large and capable models~\cite{Mikolov2011Strategies,Bengio2013Representation,Brown2020Language}, which can eventually be used for particular downstream applications~\cite{Qiu2020Pre-trained,Otter2021ASurvey}. 
However, a number of concerns have been voiced regarding the true capabilities~\cite{Bender2020Climbing}, availability, biases, and risks~\cite{Bender2021OnTheDangers}, miscalibration~\cite{Desai2020Calibration,Kong2020Calibrated}, and performance testing~\cite{Ribeiro2020Beyond} of such modern models. 
In order to improve users' understanding and eventually trust~\cite{Lee2004Trust,Hoffman2013Trust} for computational models while making use of the advantages of human perception and cognition, researchers and practitioners in NLP might benefit from the knowledge and methods developed in the information visualization~\cite{Card1999Readings} and visual analytics~\cite{Keim2010Mastering} communities, which have been traditionally open towards interdisciplinary collaborations and domain applications~\cite{Fekete2008TheValue,vanWijk2006Bridging,Sedlmair2012Design}.  

The purpose of visual text analytics (VTA) approaches in most cases is to support the users in accomplishing their tasks and achieving their goals that involve text data, computational methods including NLP/ML, and visual/interactive techniques (i.e., text visualization)~\cite{Kucher2015Text,Liu2019Bridging}. 
Examples of such approaches include systems for retrieval and exploration of patients' medical histories for physicians~\cite{Sultanum2019Doccurate} or investigation of news archives for historians~\cite{Handler2022ClioQuery}. 
In order to ensure that the proposed VTA tool indeed supports the users in achieving their goals and behaves in a reliable, predictable manner, it is imperative for designers of VTA solutions to consider the respective evaluation/validation concerns~\cite{Carpendale2008Evaluating,Lam2012Empirical,Purchase2012Experimental,Isenberg2013ASystematic,Elmqvist2015Patterns}. 
The choice of appropriate evaluation techniques and experimental designs is not a trivial issue, though, and it presents a number of open challenges.
Some of such challenges include the trade-off between the study precision, realism, and generalizability~\cite{Carpendale2008Evaluating}, experimental data collection and analysis~\cite{Purchase2012Experimental}, scope and focus of the evaluation~\cite{Lam2012Empirical}, expected reporting rigor~\cite{Isenberg2013ASystematic}, and applicability of typical evaluation patterns~\cite{Elmqvist2015Patterns}, among others.

The need for appropriate evaluation and experimental design strategies was in fact an important dimension of the discussions at a recent Dagstuhl seminar on VTA (\url{https://www.dagstuhl.de/22191}), which included 28 researchers and practitioners from both the NLP and visualization communities (10 and 18 participants, respectively; the overall group was also relatively well balanced with 11 females and 17 males, including 11 junior and 17 senior participants in terms of expertise). 
This diverse group has identified and discussed a number of VTA aspects and challenges in smaller subgroups. 
These topics included, among others, the diversity of data sources and applications of VTA, uncertainty and bias in VTA, NLP model interpretation, and visual representations for text data.

\begin{table*}[t!]
  \caption{The main groups of challenges and further topics related to evaluation and experimental design for VTA identified within our discussion group consisting of visualization and NLP researchers.}
  \label{tab:vta-evaluation-discussion-topics}
  \scriptsize%
  \centering%
  \renewcommand{\arraystretch}{1.4}%
  \linespread{0.95}\selectfont%
  \begin{tabu}{l X[l]}
  \toprule
  \textbf{High-level challenge group} & \textbf{Examples of specific concerns}\\
  \midrule
  Data ambiguity & Language/text data ambiguity; lack of single ground truth; making use of interactive visualization to reveal issues with raw and annotated data; concept drift implications for data and models\\ 
  Experimental design & Taxonomies/categorizations of users, data, NLP tasks \& scales, and respective interactive visualization evaluation methods; representation and evaluation issues specific to text data; evaluation for tools with multiple views and interactions \\
  User trust & Ensuring user trust for NLP models and complete VTA solutions; going beyond time and error measurements on VTA evaluation; considering interpretability and explainability of models and decisions for trust in VTA; measuring the effects of model errors/noise on trust\\
  VTA evaluation ``big picture'' & Systematic analysis of previous VTA evaluation efforts; guidelines for conducting and reporting VTA evaluations; repository of standard experimental designs for VTA; reproducibility concerns for case and user studies; long-term VTA adoption studies in real-world scenarios; retrofitting human-centered evaluation approaches for NLP models and tools \\
  \bottomrule
  \end{tabu}%
  \vspace{-1.65mm}
\end{table*}

The diversity of the methods and traditions existing across these two disciplines has implications for the overall design and validation process of VTA solutions. 
In this regard, the need to establish the common ground between the NLP and visualization community representatives in order to make progress in VTA research was evident, as even the term \cit{evaluation} leads to different associations and expectations in these disciplines, while the efforts from both sides are actually required.
For example, the concerns regarding the performance of the underlying computational models (e.g., a review text classifier) must be complemented with human-centered concerns driven by the expectations and reactions of various actors involved with such VTA solutions, including the target end users (e.g., marketing managers), but also data annotators and experts in charge of the NLP/ML models, for instance. 
The existing bodies of knowledge on data annotation~\cite{Artstein2008InterCoder}, performance evaluation of computational models for NLP~\cite{Sokolova2009Performance}, design and evaluation in information visualization and visual analytics~\cite{Munzner2009ANested,Meyer2012TheFourLevel,Sedlmair2012Design,Isenberg2013ASystematic}, and specifically design and evaluation of text visualization and VTA techniques~\cite{Kucher2015Text,Shamim2015Evaluation,Liu2019Bridging,Alharbi2019SoS} provide us with initial pointers and particular recipes. 
However, we would argue that truly interdisciplinary efforts are necessary to advance the understanding and practice of evaluation of VTA approaches, and furthermore, the results of such efforts could be beneficial for the respective communities beyond VTA-specific findings. 

In this position paper, we report on the outcomes of the discussion that culminates in a vision of four groups of open challenges in evaluation and experimental design for VTA from the interdisciplinary perspective across the NLP and visualization communities. 
The authors of this manuscript formed a discussion group focusing on the issues of evaluation and experimental design at the seminar. 
Our group featured an equal number of participants with a background in visualization and NLP (4+4), which helped us with the identification of concerns and concepts related to all parts of the (visual) text analytics workflow that affect the choice of evaluation methods. 
More specifically, our prior experiences included the work on NLP problems such as disagreement and errors in data annotation~\cite{Plank2014Linguistically,Uma2021Learning}, semantic role labeling~\cite{Daza2020X-SRL}, and stance analysis~\cite{Simaki2020Getting}, among others. 
With respect to text visualization and VTA, our experiences included works on clinical text visualization~\cite{Sultanum2019More,Sultanum2019Doccurate}, VTA support for data annotation and active learning for stance analysis~\cite{Kucher2017Active}, investigations of news archives~\cite{Handler2022ClioQuery}, online reviews~\cite{Jasim2022Supporting}, historical~\cite{Boukhelifa2015Supporting} and scientific~\cite{Caillou2021Cartolabe} document collections, as well as meta-analyses of the text visualization field~\cite{Kucher2015Text,Kucher2018TheState,Baumer2022OfCourse}.
Furthermore, several of the group participants shared prior interdisciplinary collaboration experiences on topic~\cite{Skeppstedt2018Topics2Themes} and stance~\cite{Simaki2020Annotating} analysis problems. 

The discussion of our group at the seminar started with a short round of introductions and interests with respect to the group topics (some of which were shaped by our prior experiences mentioned above), and then we proceeded with a brainstorming session using sticky notes from individual participants. 
We analyzed the contents of the collected notes and organized them into several categories listed in \autoref{tab:vta-evaluation-discussion-topics}. 
The following major topics were then discussed in smaller subgroups and eventually with other group participants:
\begin{itemize}
\item How should we account for the \emph{lack of single ground truth in the complex and ambiguous real-world text data}? How can interactive visualization complement the traditional methodologies of data labeling and metric-based performance evaluation used in computational linguistics?
\item What are the concerns that affect the \emph{experimental design for text visualization and interaction techniques} in VTA approaches, taking the specifics of the respective NLP techniques into account?
\item Moving beyond the design and evaluation of performance and usability, how can we ensure that the validation process accounts for the \emph{users' trust in the models} used by VTA solutions?
\item What is the existing body of knowledge of evaluation in VTA approaches? Which concerns should be taken into account when conducting \emph{systematic analyses} of the existing VTA evaluation studies? How can \emph{guidelines for future studies} be shaped?
\end{itemize}
\noindent The rest of this paper is organized according to these four topics. 
In \autoref{sec:data-ambiguity}, we highlight the issues of the traditional performance-based evaluation of NLP models (according to the usual use of the term ``evaluation'' in the NLP community) and highlight the opportunities for interactive/visual approaches to contribute to the respective activities carried out by NLP researchers and practitioners. 
Afterwards, in \autoref{sec:experimental-design}, we propose the dimensions of experimental design for human-centered modules of VTA solutions that follow from a discussion among NLP and visualization researchers in our discussion group. 
We argue for the importance of ensuring user trust for VTA systems as part of evaluation efforts in \autoref{sec:trust} and highlight the specific issues of trust assessment that should be considered. 
These and further challenges lead us to \autoref{sec:systematic-analysis}, where we argue for the need to review the ``big picture'' of VTA evaluation from an interdisciplinary perspective (including NLP, visualization, and further concerns) and propose first steps towards such a review. 
Finally, the paper is concluded with \autoref{sec:conclusion}.

\section{Concerns of text data ambiguity and lack of single ground truth}
\label{sec:data-ambiguity}

While the visualization community primarily associates the term ``evaluation'' with human-centered validation approaches for particular techniques or tools, it is important to establish that the NLP community has different associations in this regard. 
Evaluation of NLP models is typically carried out by comparing the automatic classification performed by the model to a gold standard based on manual annotations~\cite{Basile2021It}. 
That entails that the evaluation of any system based on NLP models is, in part, dependent on a gold standard. 
The creation of a gold standard for text data is not always a straightforward task: we typically assume that there is a single ground truth, and that an objective gold standard expressing this single ground truth can be created.
In some cases, this assumption is, however, an oversimplification.
Text data is rich and ambiguous, and there is no single canonical data representation or universally accepted interpretation \cite{Poesio2005Reliability,Plank2014Linguistically,Aroyo2015Truth,Pavlick2019Inherent,Basile2021It,Baumer2022OfCourse}. 
Ambiguity is a pervasive language phenomenon that can be found at all levels of linguistic analysis, from the phonological (\emph{too} vs \emph{two}) and the grammatical level (\emph{run} as a noun vs \emph{run} as a verb) to the semantic and conceptual level (as in the sentence \emph{\cit{I cannot find the mouse anywhere in the room}} that can have different interpretations in context depending on whether \emph{mouse} refers to the animal or the computer device).

Ambiguity in language may lead to different interpretations of the text by the annotators, which as a result affect their final decisions. This is a frequent phenomenon especially in discourse annotation, as some relations such as \emph{cause and effect} or \emph{contrariety} are explicitly signalled in discourse by a connector/cue phrase (e.g., \emph{because}, \emph{but}) while other may not contain any linguistic items to signpost a more implicit relation~\cite{Hoek2017Evaluating}. This may affect the annotator's understanding of the text and thus their decision.
But even in cases where annotators have a similar (or even identical) understanding of the text and they rely on a detailed, restrictive and well-crafted annotation protocol, their decisions may still differ at the end due to factors related to their perceptual system, pragmatics, and their background.

The following example illustrates plausible disagreement between two annotators (A and B) in a stance detection task:

\begin{quote}
Instance: \emph{\cit{even when our passions run high, our compassion must run deep.}} --- Kevin Deyoung 

Labels: \emph{necessity} (Annotator A), \emph{source of knowledge} (Annotator B)
\end{quote}

\noindent In the example above, taken from a stance annotation task on Twitter as discussed in our previous work~\cite{Simaki2020Getting}, Annotator A labeled the tweet with the \emph{necessity} tag, probably due to the presence of the  auxiliary verb \emph{must} that denotes necessity, while Annotator B labeled the tweet as \emph{source of knowledge} relying on the person's name (Kevin Deyoung) as the source of the quoted text.
Here, the annotators have a different perception of the overall stance expressed in this tweet, which one is the dominant one, and they probably do not share the same idea regarding the purpose of the tweet's author. Such disagreements are observed when refined concepts, such as stance, are to be identified, grouped and annotated into different categories~\cite{Simaki2020Annotating}. 

In some contexts, such as communication  or language variation studies, the different perceptions of readers over a text due to language ambiguity may be an interesting phenomenon to explore and derive insights~\cite{Piantadosi2012Communicative}. From the NLP perspective though, in the cases where the divergence between the annotators’ decisions is high, the task needs to be more closely specified (and/or simplified) and the annotation guidelines further refined, in order to reach an acceptable level of inter-annotator agreement (IAA)~\cite{Artstein2008InterCoder}. 
Traditionally, high IAA implies higher data quality (in terms of reproducibility of data labeling). 
But due to the nature of language itself and the various meanings that exist, assuming high agreement (and by consequence, a single ground truth) is at best an idealization~\cite{Uma2021Learning}. This assumption of a ground truth neglects genuine disagreement, which can provide valuable insights about language use and interpretation.  
For some text interpretation tasks, too restrictive guidelines might even hide the fact there is no real ground truth for this particular task.  While the body of work on learning from disagreement is growing in NLP, including our own prior efforts~\cite{Uma2021Learning}, there is a lack of interdisciplinary research between the NLP and visualization communities in this regard.

\begin{figure*}[ht!]
\centering
    \includegraphics[width=0.75\linewidth]{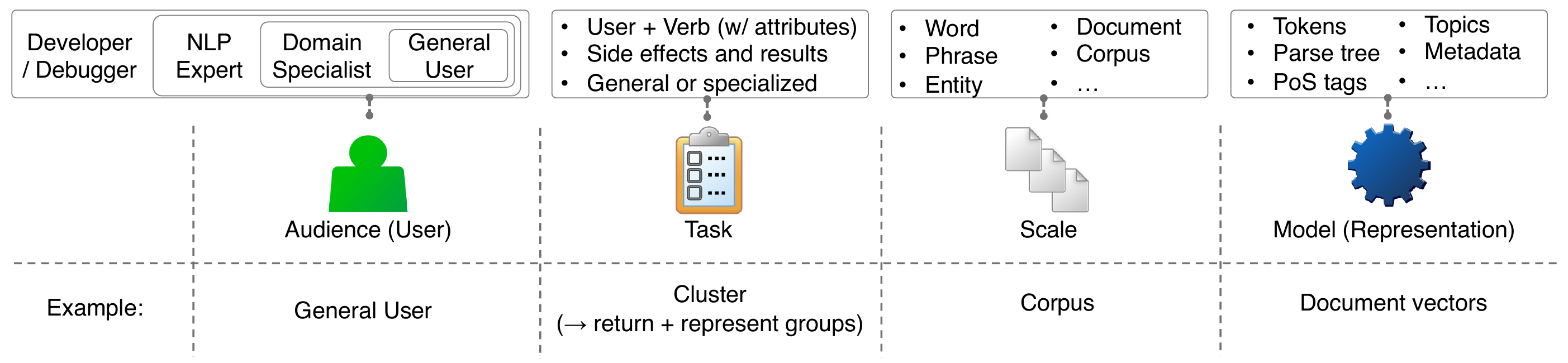}
	\caption{Dimensions of experimental design for text visualization techniques that take into account the specifics of the underlying NLP tasks and representations.%
	}%
	\label{fig:vta-experimental-design-dimensions}%
\end{figure*}

In particular, there are many examples of visualizations that create an overview of the results of text annotations, for visualizing the co-occurrence of different annotation categories. 
Approaches range from using standard visualization techniques (e.g.\ bar charts, scatter plots, Sankey and chord diagrams, and treemaps~\cite{Kavaz2021Data}) to developing new visualization techniques specific to the text annotation task (e.g., the ALVA tool for supporting annotation and active learning for stance analysis~\cite{Kucher2017Active}, which was developed in the scope of prior collaboration between several authors of this manuscript~\cite{Simaki2020Annotating}). 
 However, user interfaces for showing the difference in annotation categories between different annotators typically focus on 1) individual annotation instances (e.g., WebAnno~\cite{Yimam2013WebAnno}), or 2) IAA statistics for the annotators, such as matrices showing the inter-annotator agreement for pairs of users (e.g., ERAS~\cite{Grosman2020Eras}). 
That is, the aim of these visualizations is not to provide the user with an overview of the differences in annotation choices. 
This is natural, as these interfaces typically are created for curators who use them for annotation adjudication~\cite{Biemann2017Collaborative}, i.e., merging each conflicting annotation into the one correct annotation, rather than exploring the annotated data set in order to gain insights into the reasons for disagreement. For instance, insights into to what extent there is a single ground truth for the annotation task.

We believe that VTA provides opportunities for mutual benefits of research within NLP and text visualization to provide insights into the nature of disagreement. 
In addition to visualizing disagreement between annotators in terms of labeling divergences, there are additional factors that contribute to disagreement and could be visualized. 
For instance, the annotators could be given the opportunity to not only annotate, but also specify the level of uncertainty for the annotation, or to opt out of annotating an instance due to it being too difficult. 
Also the time it takes for the annotators to make an annotation decision could be gathered as an information source to indicate the difficulty of the task.

In order to make sense of all this information on the annotators and the annotation process, and decide in what direction to steer the annotators (to make the task more restrictive, or to accept that there is is no single ground truth), the following information provided from the annotation process could be visualized: 

\begin{itemize}
\item For which annotation cases do the annotators agree?
\item For which annotation cases do the annotators disagree?
\item How can you...
    \begin{itemize}
	\item First provide an overview to what extent they agree/disagree;
	\item Provide an overview of the cases with agreement/disagreement;
	\item Make it possible to zoom in to different cases of agreement and disagreement;
	\item Let the user zoom in on groups of cases of disagreements, and on the actual instance of disagreements;
	\item Let the user make their own classifications of the agreement and disagreement and also make it possible to visualize these classifications; and 
	\item Visualize automatically derived classifications of annotator agreement and disagreement.
    \end{itemize}
\end{itemize}

\noindent It might also be interesting to compare different annotation tasks, or the same annotation task on different corpora for their level of agreement/disagreement. 
A visualization might provide insights and a quick overview into the difference between different corpora, and thereby into the level of difficulty automating a classification task.

Most importantly, giving the possibility to let the user understand agreement/disagreement between annotators on different levels might help one get insight into whether the annotation task should be more clearly specified, or if there is no single ground truth for the text annotation task that has been given to the annotators.

\section{Concerns for experimental design: Audience, tasks, scale, models}
\label{sec:experimental-design}

Moving from the data-driven concerns towards the issues of validation of text visualization approaches, we consider that functionality of 
an NLP system and the respective (interactive) visualizations should be evaluated in light of at least the following four dimensions (see \autoref{fig:vta-experimental-design-dimensions}): target \emph{audience} (who?), list of \emph{tasks} to support (why?), \emph{scales} to consider in the tasks (linguistic objects of interest), and \emph{models} supporting the tasks at the given scales (how?).

\textbf{Audience}: we start with the users who define what they want to perform in the system (goals). 
We can define the following User Groups (from larger to smaller in terms of the system knowledge and expected functionality): 1) \emph{Debugger} or \emph{Developer} of the system, 2) \emph{Linguist} or \emph{NLP Expert} in charge of the modeling, 3) \emph{Domain Specialist} using the system with no particular knowledge of NLP or CS (e.g., a physician or a historian), and 4) \emph{General User}, where each member of a group has the ability to perform the tasks of their own group and the smaller groups contained within (cf. \autoref{fig:vta-experimental-design-dimensions}). 
Several examples of our own published studies demonstrated the use of VTA approaches with public healthcare~\cite{Skeppstedt2018Topics2Themes}, clinical~\cite{Sultanum2019Doccurate,Sultanum2019More}, and civic~\cite{Baumer2018Interpretive,Baumer2022OfCourse} text data sets, for instance.  
Considering the real-world implications of the use of such approaches, the importance of carefully considering the intended audience and their goals for design and evaluation of visual text analytics cannot be overstated. 

\textbf{Tasks}: we then define the tasks that will achieve each of the users’ goals. 
A task is the combination of a User + Verb (and attributes); it can lead to a side effect (change in the visualization) or to a result (list of items). 
For example: \emph{overview} (depends on the visual representation, no parameters), \emph{query}, \emph{search} (return a list of items), \emph{cluster} (visualize groups and return a list of groups), etc. 
Some prototypical tasks are:
\begin{itemize}
\item Search documents containing given text snippets;
\item Search sentences containing two adjectives;
\item List documents similar to document X; or
\item Group sentences containing keywords.
\end{itemize}
\noindent \emph{Specialized} tasks can be domain-specific, but also specific to the linguistic model (aimed at the NLP specialist) or to the programmer, allowing the respective user to understand the internal data structure of the NLP and visualizations. 
Such tasks are not useful for the domain expert or to a lay user. 
Further potential tasks can be identified in the previous works on the more general task taxonomies in information visualization and visual analytics~\cite{Amar2005Knowledge,Schulz2013ADesign,Brehmer2013AMulti-Level,Kerracher2017Constructing,Lam2018Bridging} as well as the sources focusing on text visualization and visual text analytics~\cite{Wanner2014State,Tofiloski2015ATaxonomy,Shamim2015Evaluation,Cao2016Introduction,Chen2017Social,Jaenicke2017Visual,Liu2019Bridging,Alharbi2019SoS}, including the studies contributed by the authors of this manuscript~\cite{Kucher2015Text,Kucher2018TheState}, among others.

\textbf{Scale}: each (or almost each) task can be applied or affect the text at a different level of granularity: \emph{word}, \emph{phrase}, \emph{entities}, \emph{tree structure} (syntax), \emph{proposition} (semantics), \emph{sentence}, \emph{section}, \emph{paragraph}, \emph{document}, \emph{corpus}, or \emph{stream}~\cite{Wanner2014State}, for instance. 

\textbf{Models}: layers of linguistic information are encoded and structured differently: \emph{parse tree}, \emph{part-of-speech tags}, XML-encoded text (\emph{metadata}), \emph{stand-off annotation} (e.g., relevant spans in text), \emph{topics}, etc.
The model influences the tasks in the sense that it allows some tasks to be done more or less accurately, and with diverse levels of uncertainty, quality, errors, and artifacts. The models also constrain the scale of possible tasks and affect the choice of suitable visual encodings and interaction techniques.

\begin{figure*}[ht!]
\centering
    \includegraphics[width=0.995\linewidth]{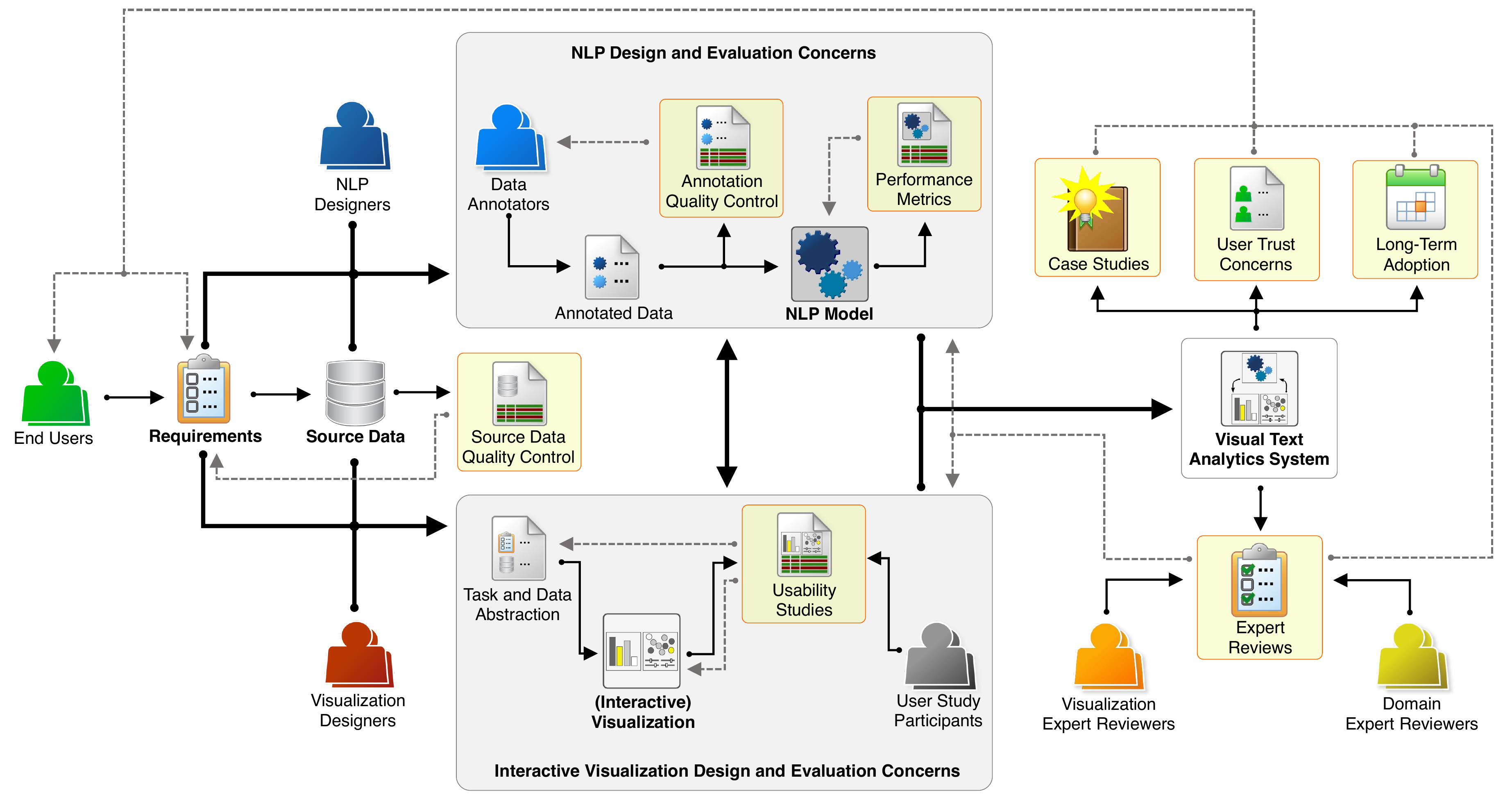}
	\caption{Sketch of a visual text analytics design and evaluation workflow with explicit identification of validation concerns which can guide our future efforts for systematic analysis and preparation of guidelines. %
	The nodes with bold font labels indicate the main components and artifacts leading to the implementation of VTA solutions; the nodes highlighted with yellow background indicate quality control and validation activities; 
	the grey shaded boxes represent the areas focusing on the particular NLP and visualization concerns; 
	black edges indicate interactions and effects; thick black edges indicate the main interactions and effects leading to the implementation of VTA solutions; and grey dashed edges indicate feedback effects resulting from quality control and validation activities.%
	}%
	\label{fig:vta-pipeline-design-evaluation-concerns}%
\end{figure*}

\section{User trust in automation in the context of visual text analytics evaluation}
\label{sec:trust}

Given the complex, multifaceted, and potentially ambiguous nature of text, there is a wide error range in NLP outcomes, as we cannot expect NLP systems to work perfectly in every circumstance, e.g., for each task and for each data set. 
This is something users must take into account when leveraging text visual analytics systems for a particular problem, and is reflected in the extent users trust automated outcomes and take them into account. 
User trust is an evolving process~\cite{Hoffman2013Trust}, and it reflects perceived system performance over use and the risk associated with decision-making. 
For example, spelling and grammar correction in documents is a relatively mature problem with low impact for errors, and something users are generally comfortable delegating to automation. 
On the other hand, automatic summarization of patient records for clinical decision making is an open research problem with significant consequences for wrong medical choices~\cite{Pivovarov2015Automated}. 

Trust is a crucial factor when measuring overall task performance and user experience. 
Misplaced trust leads to misguided decision-making, when relying too much on wrong assessments (i.e., overtrust), generating user overhead in verifying largely accurate outcomes (i.e., undertrust), and in some cases leads users to completely ignoring the results due to lack of results and preferring the manual sensemaking process~\cite{Muir1987Trust}, as observed in our several prior studies~\cite{Sultanum2019Doccurate,Sultanum2019More,Handler2022ClioQuery,Jasim2022Supporting}, for instance.
As such, there will always be a need for a system that uses NLP to communicate to what extent the user can trust the automatic output of the system~\cite{Sacha2015Role}. 
In visual text analytics, this is often enabled by providing access to original text sources, allowing users to calibrate their trust. 
Allowing users to fix errors is another important consideration~\cite{DeArteaga2020Case}, but propagating corrections across a dataset is not always a straightforward task. 
In the end, while guidelines have been proposed to communicate uncertainty and foster trust in visual (text) analytics systems~\cite{Lee2004Trust,Chuang2012Interpretation,Chatzimparmpas2020TheState,Sacha2015Role}, designing for trust is not always seen as a first class concern.

Apart from design considerations to foster adequate trust, from an evaluation perspective, it is also important to consider how trust can be measured, and whether it matches NLP capabilities~\cite{Kohn2021Measurement}. 
For example, the study by Dasgupta et al.~\cite{Dasgupta2017Familiarity} operationalizes user trust as a degree of perceived confidence in the analysis results; the self-reported confidence can then be compared across various user tasks, for instance. 
While the methodology of that study might be adapted to at least some VTA evaluation efforts, in general there are many challenging components to measuring adequate trust within a visual text analytics context.
It entails 1) an understanding of the limitation areas of NLP algorithms within a particular application, 2) assessing user perception of the limits of automated outcomes, 3) assessing how well user perception matches real capabilities of automation, and 4) how trust levels evolve over time. 
While there is research in understanding and designing for trust, strategies for measuring trust are still not widely adopted in the evaluation of visual text analytics systems. 
Moving forward, we posit that the community should consider developing and fostering use of standardized and easily applicable metrics and instruments (e.g., questionnaires~\cite{Kohn2021Measurement}) for measuring trust in this specific context, which should be applicable at different stages of use (i.e., measurable over a period of time).


\section{Big picture: Systematic analysis and guidelines for evaluation of visual text analytics}
\label{sec:systematic-analysis}

One of the challenges of evaluating visual text analytics approaches is related to the breadth and variety of concerns regarding computational and interactive, human-centered aspects of such approaches and tools. 
The efforts focusing on systematic analysis, categorization, and formulation of guidelines for evaluation (cf. the work by Isenberg et al.~\cite{Isenberg2013ASystematic}) would thus be beneficial for researchers, practitioners, and end users of such visual text analytics tools. 
Here, the topics and dimensions would include an interdisciplinary analysis/view and include both NLP/ML-focused aspects (e.g., the methods and measurements of intra- and inter-annotator agreement, or the measurements of NLP model performance for a given task beyond standard ML metrics such as accuracy or F$_1$-score) as well as human-centered aspects (e.g., the methods of usability measurement and long-term adoption observation). 
While the complexity and variety of visual text analytics problems would most likely not allow us to recommend a single prescribed method for validation, the outcomes of the work on systematic analysis would result in a collection of guidelines that would benefit the research efforts and applications.

One possible way to look at the issues of VTA evaluation from an interdisciplinary perspective (and taking the issues and opportunities for cross-/interdisciplinary effects and benefits, as discussed in the previous sections) would be to consider a sketch of a possible complete (visual) text analytics design and evaluation workflow. 
Inspired by the well-known visual analytics knowledge generation model by Sacha et al.~\cite{Sacha2014Knowledge}, we consider the relations between the data, NLP model, visualization, users, researchers/designers from both NLP and visualization communities, and the respective design and evaluation concerns.  
In the rest of this section, we focus on the sketch depicted in \autoref{fig:vta-pipeline-design-evaluation-concerns} and the respective actors, activities, and artifacts, starting with the first steps towards the design of VTA solutions from scratch (the left part of the figure):

\textbf{End users} and \textbf{requirements}: while listed as \cit{end} users, it is typically the intended goals and requirements of users that guide the design of visual text analytics approaches, for instance, an expert in bioinformatics might be interested in an interactive solution for exploring a large collection of research literature in their field and finding related work based on their interests, or a user of some social media platform might be interested in checking their own personal data over time with respect to sentiment and topics~\cite{Chen2017Social}, or a product manager might be interested in investigating the opinions expressed in customer reviews. 
From these examples as well as the considerations given above in \autoref{sec:experimental-design}, we can establish that the level of expertise with respect to NLP and visualization can be very different from scenario to scenario. 
The previous work on this part of the VTA design and evaluation workflow includes, for instance, the study focusing on domain experts as the target users by Wong et al.~\cite{Wong2018Towards}, previously presented at BELIV, and the discussions of pre-design and discovery activities for visualization design and evaluation~\cite{Sedlmair2012Design,Lam2012Empirical}. 

\textbf{Data}: this includes all the different types of language data, either from written and/or oral discourse, in one or several languages~\cite{Ponti2019Modeling,Daza2020X-SRL}. 
Depending on the nature of the data and its sources, different methods and techniques are used to extract and process it in order to bring it to the desired shape for the different NLP and visualization tasks. 
In many cases, complementary material is included in the data set to address ambiguity at different levels, as for instance in data from sign language (videos with gestures) and texts transcribed from oral discourse (audio files). 
The different genres, content, topics, and communicative situations/purposes are also important characteristics to be identified, as discussed in \autoref{sec:data-ambiguity}. 
Furthermore, for complete VTA solutions, the raw data may include not only language/text information, but also further data types not directly addressed by NLP methods, for example, user metadata from social media platforms~\cite{Chen2017Social}. 

\textbf{Data quality control}: it is worth noting that before the specific design decisions for NLP and visualization are made, the raw data associated with the intended VTA application should be assessed, as the issues of missing, duplicate, or ambiguous records may have to be identified and addressed in some way. 
The previous work on data quality assessment can be considered here~\cite{Pipino2002Data}, including the implications for NLP/ML~\cite{Garcia2010Pattern} and potential data augmentation techniques for NLP~\cite{Feng2021Survey}, among others. 

We are now proceeding from the left to the central part of the workflow figure.

\textbf{NLP designers} and \textbf{visualization designers}: we have added the actors responsible for the overall NLP and visualization components in the VTA design and evaluation workflow sketch in \autoref{fig:vta-pipeline-design-evaluation-concerns}, but please notice that the same individuals might be \cit{wearing a different hat} and play further roles within the design and evaluation process, e.g., as data annotators or eventual end users. 
We discuss the NLP components next (the bordered block in the upper central part of the workflow sketch), and then address the visualization components (the lower central part). 

\textbf{Annotators}: an annotator is the person who is asked to read, assess, and label a given text according to specific instructions. 
Depending on the purpose of the study, it can be a domain expert (e.g., a linguist observing dialectal characteristics in language data collected from a specific region, or a historian assessing the relevance/trustworthiness of materials and sources regarding the French revolution) who has a background and knowledge relevant to the study. 
In many cases, the domain expert is also the one who designs the study and the annotation task, and assesses the results. 
In other cases, the annotator can be randomly selected and they can either belong to a specific target group (e.g., native speakers of English annotating the quality of texts written by learners of English), or just people interested in participating in a coding task (e.g., via crowdsourcing~\cite{Snow2008Cheap}).

\textbf{Annotation process} and \textbf{annotation quality control}: as discussed in \autoref{sec:data-ambiguity}, the process of attributing labels to text may differ significantly depending on the nature of the data and the purpose of the study. 
However, some general rules apply for the implementation of the study. 
A protocol that guides the annotator is drawn, in which the research expert define the concepts involved and describe the various steps that need to be followed. 
Based on the annotation task, annotators are selected and a pilot round is conducted. 
The results of this round are discussed with the research expert, and arisen issues are discussed in order to be addressed in the upcoming steps of the study. 
This can lead to refinements and/or changes in the annotation protocol. 
After this, the actual annotation round takes place. 
The annotation outcomes can be statistically evaluated, usually by using metrics such as Cohen's kappa~\cite{Cohen1960Coefficient} that calculate the agreement between the annotation decision of two or more annotators (IAA). 
Opinions regarding the most suitable metrics for the IAA calculation and the acceptable level of this agreement vary, and researchers have not reached a consensus towards a common path to evaluate the quality the annotation results~\cite{Artstein2008InterCoder,Gwet2002Kappa,Lombard2002Content}, which provides opportunities for further contributions.

\textbf{Computational (NLP) models for text analysis}: without going into excessive detail on the choice of particular algorithms and activities such as hyperparameter search, which are well-covered by the related work in NLP and ML, we should mention the issues of over-relying on pre-trained models~\cite{Bender2020Climbing,Bender2021OnTheDangers} and their trustworthiness, which might require additional calibration~\cite{Kong2020Calibrated,Desai2020Calibration}. 
Such issues have implications not only for the NLP model itself, but also other components of the VTA design and evaluation workflow.

\textbf{Model evaluation methods and results}: the traditional approach for assessing performance of NLP models relies on quantitative estimation of the model error and performance metrics calculated in relation to the ground truth data, such as accuracy, precision, recall, F$_1$-score, etc.~\cite{Sokolova2009Performance}. 
However, not only numbers of hits and misses should be taken into account, but also the respective task, audience, etc., as discussed in \autoref{sec:experimental-design}. 
The excessive reliance on benchmark performance evaluation of NLP models has also drawn criticism due to the eventual detriment of such properties as model size, energy efficiency, and practical utility for real-world language data~\cite{Bowman2021What,Ethayarajh202Utility,Schlangen2021Targeting}.  
In this regard, we believe that there is an opportunity for the NLP community to benefit from collaboration with the visualization community and make use of some of the human-centered evaluation considerations. 
Furthermore, the need for interactive approaches to allow NLP experts to prepare and investigate additional tests for their models in an efficient and flexible way has been recognized by the community~\cite{Ribeiro2020Beyond}. 

We now switch to the visualization design and evaluation concerns (the lower central part in \autoref{fig:vta-pipeline-design-evaluation-concerns}).

\textbf{Task and data abstractions}: this is a more familiar territory for researchers in visualization, and the existing work on the respective methodologies~\cite{Amar2005Knowledge,Munzner2009ANested,Meyer2012TheFourLevel,Sedlmair2012Design,Brehmer2013AMulti-Level,Kerracher2017Constructing,Lam2018Bridging} will definitely provide the basis for systematic analyses of VTA evaluation efforts. 
At the same time, we should take the concerns specific to language/text data and NLP models into account, too~\cite{Tofiloski2015ATaxonomy}, as discussed in \autoref{sec:experimental-design}. 

\textbf{Interactive visual representations}: to continue the previous point, we could comment on one representation that is widely used for text visualization tasks (and is often explicitly requested by end users), while raising issues with visualization researchers---word clouds~\cite{Viegas2008Tag} and their varieties. 
The various design issues and recommendations for application of this technique have been discussed in the literature~\cite{Alexander2018Perceptual,Felix2018Taking,Hearst2020AnEvaluation}, and it is an example of VTA design and evaluation concerns that must be considered in context of the target audience, tasks, and underlying representations (cf. \autoref{sec:experimental-design}). 

\textbf{Usability studies}: while the traditional human-computer interaction approach for empirically assessing usability of (interactive) visualization techniques typically relies on task completion time and error rate measurement in laboratory settings~\cite{Frokjaer2000Measuring}, the visualization community (including the BELIV workshop) has long argued that this is not sufficient. 
Even considering the text visualization essentially on its own at this point (without the NLP model and synergy concerns), we can make use of the rich body of knowledge existing in the community in this regard~\cite{Purchase2012Experimental,Lam2012Empirical,Isenberg2013ASystematic,Elmqvist2015Patterns}, including the work focusing on text visualization in particular~\cite{Baumer2018Interpretive} and even making use of traditional approaches for analyzing text data from the humanities~\cite{Bares2020Using}, for instance. 

\textbf{User study participants}: a particular role included in our workflow sketch in \autoref{fig:vta-pipeline-design-evaluation-concerns} is of participants invited to the user studies of (interactive) text visualization approaches. 
It is important to remember of the audience and task considerations here (cf. \autoref{sec:experimental-design}), as undergraduate students or members of the general public may or may not be the suitable audience for particular studies, posing a risk for their validity. 
Furthermore, while we mentioned the existing efforts on making use of crowdsourcing for data annotation on the computational side of VTA~\cite{Snow2008Cheap} above, there are also discussions of feasibility and concerns of such activities for user studies~\cite{Archambault2017Evaluation}, which should also be taken into account for VTA evaluation.

After focusing on the NLP and visualization components on their own, we finally reach the stage of complete VTA solutions (see the right part of \autoref{fig:vta-pipeline-design-evaluation-concerns}), with the expectation of helping users achieving better results, knowledge, and particular insights through the visual analytics process~\cite{Keim2010Mastering}.

\textbf{Visual text analytics solutions}: the extra capacity and flexibility of such approaches might come at the cost of overcomplicated solutions relying on dozens of NLP models and dozens of interactive views; 
the issue is, once again, how to ensure that such solutions work well for the intended audience, tasks, and data. 
The problem of evaluating visual analytics systems in comparison to individual views or interactions has been raised for quite some time in the visualization community, and we would argue that it is still an open challenge. 
However, the relevant considerations, e.g., for holistic vs reductionist views on evaluation~\cite{Correll2014Navigating} or evaluation of interactive visual machine learning systems~\cite{Boukhelifa2020Challenges}, should definitely be included in the future systematic analysis efforts for VTA evaluation.

\textbf{Expert reviews}: some of the existing evaluation methods that could be applied to the complete VTA solutions rather than individual underlying techniques include reviews by the experts. 
Here, we could specifically refer to the work focusing on the feedback from experts in visualization and human-computer interaction~\cite{Tory2005Evaluating}, including not only interviews, but also structured questionnaires such as ICE-T with the focus on value added by visualization~\cite{Wall2019AHeuristic}; and the work focusing on the involvement of domain experts~\cite{Elmqvist2015Patterns}. 
One comment that we should add in this regard is that the recommendations (or even requirements) for reporting the results of such reviews should cover the explicit statement on the overlap or other conflicts of interest between the experts and the authors of respective research studies (which might not always be evident, e.g., during the double-blind manuscript review process).

\textbf{Case studies}: we are now reaching the final part of the workflow sketch proposed in \autoref{fig:vta-pipeline-design-evaluation-concerns}, where the complete VTA solutions are applied by the intended end users. 
The outcomes of such applications with real-world data, reported as case studies, provide another opportunity for evaluation~\cite{Lam2012Empirical,Isenberg2013ASystematic,Elmqvist2015Patterns}, and the particular studies should be included in the systematic analysis efforts for VTA evaluation. 
One additional note to mention here is related to the reporting of case study data, steps, and results, which should contribute to the reproducibility of the results by reviewers and eventual readers of the respective studies~\cite{Fekete2020Exploring}. 

\textbf{User trust concerns}: as discussed in \autoref{sec:trust}, the issues of user trust~\cite{Lee2004Trust,Chuang2012Interpretation, Sacha2015Role} permeate the VTA design and evaluation process (and the positioning of the respective node in \autoref{fig:vta-pipeline-design-evaluation-concerns} can be seen as an oversimplification, for instance). 
The existing work on trust in automation, visual analytics for enhancing trust~\cite{Chatzimparmpas2020TheState}, evaluation of the respective concerns in visualization and visual analytics~\cite{Dasgupta2017Familiarity}, and specific work on text visualization and VTA should be considered for the future systematic analysis efforts; furthermore, we envision a number of opportunities for future work in this direction.

\textbf{Long-term adoption}: finally, we should consider longitudinal studies or other potential approaches to gauge adoption of VTA solutions in the long term by the respective audiences. 
Such evidence, including observations and studies conducted outside of the academic environment, for instance, in commercial~\cite{Sedlmair2010Evaluating,Russell2016Simple} and public~\cite{Baumer2022OfCourse} settings, will be extremely valuable for both visualization and NLP communities. 

\section{Conclusion}
\label{sec:conclusion}
In this position paper, we have taken an interdisciplinary perspective on evaluation of visual text analytics approaches based on the existing work and our own prior knowledge in NLP and visualization. 
Our main message is that there is definitely interest and opportunity to benefit from collaborations in the scope of VTA evaluation research. 
We have discussed the issues of data ambiguity and lack of single ground truth which are usual for real-world language and text data; the important experimental design dimensions to consider for evaluation of particular (interactive) text visualization techniques; the importance of user trust issues for VTA design and evaluation; and finally, the need to conduct further systematic analyses of VTA evaluation efforts. 
Our proposed sketch of a VTA design and evaluation workflow relates to a number of prior works and it has already led us to further points in our interdisciplinary group discussions, including the existing knowledge gaps within and across the disciplines. 
For instance, if we compare the typical NLP and visualization evaluation methodologies, we would notice that the visualization community often relies on time and error rates in usability studies, while the NLP community puts a lot more focus on error rates rather than time/efficiency concerns---can we learn from each other in this and further regards? 
Can we align the design and evaluation methodologies from our fields in order to leap forward together?



\acknowledgments{
This paper is a result of the discussion group work at Dagstuhl Seminar 22191 ``Visual Text Analytics''. 
The authors would like to thank the organizers and the rest of seminar participants for this opportunity and the feedback provided for our group work.
}

\bibliographystyle{abbrv-doi}

\bibliography{bibliography}
\end{document}